\documentclass[
 aip,
 amsmath,amssymb,
 reprint,
]{revtex4-1}

\usepackage[makeroom]{cancel}
\usepackage{dcolumn}% Align table columns on decimal point
\usepackage{makecell}
\usepackage[utf8]{inputenc}
\usepackage[T1]{fontenc}
\usepackage{mathptmx}
\usepackage{commath}
\usepackage{graphicx}
\usepackage{multirow}
\usepackage{amsmath}

\usepackage{soul}
\usepackage{verbatim}
\usepackage{bm}
\usepackage{outlines}
\usepackage{siunitx}
\usepackage[space]{grffile}
\usepackage[dvipsnames]{xcolor}
\usepackage{xr}
\usepackage{color}
\usepackage{tikz}
\usepackage{mathtools}
\usepackage{amssymb}

\newcommand\myshade{85}
\colorlet{mylinkcolor}{violet}
\colorlet{mycitecolor}{YellowOrange}
\colorlet{myurlcolor}{Aquamarine}
\usepackage{hyperref}
\hypersetup{
  linkcolor  = mylinkcolor!\myshade!black,
  citecolor  = mycitecolor!\myshade!black,
  urlcolor   = myurlcolor!\myshade!black,
  colorlinks = true,
}
\usepackage[capitalize]{cleveref}

\usetikzlibrary{calc, positioning}

\newcommand{\K}{\mathrm{K}}

\newcommand{\MHz}{\mathrm{MHz}}
\newcommand{\GHz}{\mathrm{GHz}}
\newcommand{\s}{\mathrm{s}}
\newcommand{\us}{\mu\mathrm{s}}
\newcommand{\ns}{\mathrm{ns}}

\newcommand{\dB}{\mathrm{dB}}

\newcommand{\Ttwostar}{T_{2}^{\ast}}

\newcommand{\ket}[1]{\left\lvert #1 \right\rangle}

\newcommand{\Phio}{\Phi_0}

\newcommand{\dd}[1]{\mathrm{d}#1}

\newcommand{\freqQ}{f_\mathrm{Q}}
\newcommand{\freqQmax}{f_{\mathrm{max}}}
\newcommand{\detuning}{\Delta \freqQ}
\newcommand{\detuningAvg}{\overline{\Delta f}_\mathrm{Q}}
\newcommand{\detuningEst}{\overline{\Delta f}_\mathrm{R}}

\newcommand{\PhiQ}{\Phi_{\mathrm{Q}}}            % Actual flux
\newcommand{\PhiR}{\Phi_{\mathrm{R}}} %Reconstructed flux

\newcommand{\PhiRmn}{\overline{\PhiR}} %Mean Reconstructed flux
\newcommand{\PhiRstd}{s_{\PhiR}} %Std Reconstructed flux

\newcommand{\hfir}{h_\mathrm{FIR}}
\newcommand{\SNR}{\mathrm{SNR}} % Signal to noise ratio
\newcommand{\aSNR}{c} % Amplitude in cryoscope SNR equation

\newcommand{\tTrunc}{\tau}
\newcommand{\tIIR}{\tau_\mathrm{IIR}}
\newcommand{\tHP}{\tau_\mathrm{HP}}

\newcommand{\phaseQ}{\varphi}

\newcommand{\phaseDiff}{\Delta\phaseQ}

\newcommand{\imp}{h}

\newcommand{\impPseudoInv}{\widetilde{h}_\mathrm{inv.}}

\newcommand{\impFilt}{h_{\mathrm{filt}}}

\newcommand{\step}{s(t)}
\newcommand{\heaviside}{u(t)}
\newcommand{\stepCorr}{s_\mathrm{corr}(t)}
\newcommand{\Tsep}{T_{\mathrm{sep}}}

\newcommand{\Ec}{E_\mathrm{C}}
\newcommand{\Ej}{E_\mathrm{J}}

\newcommand{\Vin}{V_\mathrm{in}} % Input pulse at AWG
\newcommand{\VinTrunc}{V_\mathrm{in, \tTrunc}} % Input pulse at AWG

\DeclareMathOperator\erf{erf}

\begin{document}

\title{Time-domain characterization and correction of on-chip distortion of control pulses in a quantum processor}
\newcommand{\QuTech}{\affiliation{QuTech, Delft University of Technology, P.O. Box 5046, 2600 GA Delft, The Netherlands}}
\newcommand{\Kavli}{\affiliation{Kavli Institute of Nanoscience, Delft University of Technology, P.O. Box 5046, 2600 GA Delft, The Netherlands}}
\newcommand{\ETH}{\affiliation{Department of Physics, ETH Zurich, CH-8093 Zurich, Switzerland}}
\newcommand{\ZI}{\affiliation{Zurich Instruments AG, CH-8005 Zurich, Switzerland}}

{
\makeatletter
\def\frontmatter@thefootnote{%
 \altaffilletter@sw{\@fnsymbol}{\@fnsymbol}{\csname c@\@mpfn\endcsname}%
}%
\def\@fnsymbol#1{\ensuremath{\ifcase#1\or \dagger\or \ddagger\or
   \mathsection\or \mathparagraph\or \|\or **\or \dagger\dagger
   \or \ddagger\ddagger \else\@ctrerr\fi}}

\author{M.~A.~Rol}
\QuTech\Kavli
\author{L.~Ciorciaro}\QuTech\ETH
\author{F.~K.~Malinowski}\QuTech\Kavli
\author{B.~M.~Tarasinski}\QuTech\Kavli
\author{R.~E.~Sagastizabal}\QuTech\Kavli
\author{C.~C.~Bultink}\QuTech\Kavli
\author{Y.~Salathe}\ZI
\author{N.~Haandbaek}\ZI
\author{J.~Sedivy}\ZI
\author{L.~DiCarlo}
\email[Corresponding author: ]{l.dicarlo@tudelft.nl}
\QuTech\Kavli 
\date{\today}

\begin{abstract}
We introduce Cryoscope,  a method for sampling on-chip baseband pulses used to dynamically control qubit frequency in a quantum processor.
We specifically use Cryoscope to measure the step response of the dedicated flux control lines of two-junction transmon qubits in circuit QED processors with the temporal resolution of the room-temperature arbitrary waveform generator producing the control pulses.
As a first application, we iteratively improve this step response using optimized real-time digital filters to counter the linear-dynamical distortion in the control line,  as needed for high-fidelity,  repeatable one- and two-qubit gates based on dynamical control of qubit frequency.
\end{abstract}

\maketitle

In many solid-state quantum information platforms, accurate dynamical control of qubit frequency is key to realizing single- and two-qubit gates.
Common on-chip control variables include, but are not limited to, voltage on a local gate and magnetic flux through a SQUID loop.
For example, voltage control is typically used for spin qubits~\cite{Foletti09,Medford13,Laucht15,Veldhorst15} and gatemons\cite{Larsen15,Casparis15}, while flux control is ubiquitous for transmon, flux and fluxonium superconducting qubits~\cite{Kjaergaard19}.
In most cases, the input control signal originates at an arbitrary waveform generator (AWG) operating at room temperature.
The signal suffers linear dynamical distortions as it traverses various electrical components on the control line connecting to the quantum device, most often lying at the coldest stage of a dilution refrigerator.

If uncompensated, such distortions can have detrimental effects on gate performance, affecting fidelity and even repeatablility.
A salient example is the controlled-phase (CZ) gate between two transmon qubits implemented by a baseband flux pulse~\cite{DiCarlo09} that brings the computational state $\ket{11}$ temporarily near resonance with the non-computational state $\ket{02}$.
Short-timescale distortions of the meticulously shaped flux pulse~\cite{Martinis14} can produce leakage away from the two-qubit computational subspace, leaving remnant population in $\ket{02}$.
Meanwhile, long-timescale distortions make the unitary action of a flux pulse depend on the history of flux pulses applied~\cite{Langford17, Rol19_NetZero}.
As leakage and history dependence severely limit the depth of quantum circuits that can be realized, a practical scheme for characterization and correction of pulse distortion on chip is of paramount importance.

Distortions introduced by components at room temperature (e.g., AWG bandwidth, high-pass filtering of a bias tee, skin effect in instrumentation cable) are easily characterized with a fast oscilloscope.
However, distortions introduced by components inside the refrigerator (e.g., low-pass filters, impedance mismatch, skin effect in semi-rigid coaxial cable, chip packaging~\cite{Foxen18}) are generally temperature-dependent and are thus best characterized in the cold.
Additionally, the on-chip response varies across devices and even between different qubits on the very same device.
Evidently, the ideal strategy for characterizing pulse distortion is to use the controlled qubit itself.

A traditional method to visualize the dynamical distortion of ideally square pulses is to observe the oscillations in the excited-state population (as a function of pulse amplitude and duration) when pulsing the qubit into near resonance with another exchange-coupled qubit or a continuous drive tone.
While the distortions can be gleaned from the deviation from the ideal chevron pattern~\cite{Langford17}, the inversion is challenging.
More direct methods use spectroscopy~\cite{Johnson11} and Ramsey experiments~\cite{Kelly15_thesis} to measure the qubit frequency dynamics, but only during the turn-off transients following a square pulse.
Most recently, a method combining continuous microwave and flux drives was developed to convert a transmon into a vector network analyzer~\cite{Jerger17} giving the frequency response of the flux control line,  from which it is possible to calculate the qubit frequency dynamics for a given pulse.

In this Letter, we present Cryoscope (short for cryogenic oscilloscope), an in-situ technique using the qubit to sample control pulses of arbitrary shape at the temporal resolution of the AWG.
We specifically demonstrate Cryoscope for two-junction transmon qubits, whose frequency depends quadratically (to a good approximation) on the flux through the constituent SQUID loop.
However, Cryoscope is generally applicable to any system with quadratic or higher power dependence of qubit frequency on the control variable and a sweetspot where qubit frequency is at least first-order insensitive to this variable.
As a first application, we use Cryoscope to iteratively measure the voltage-to-flux step response and apply predistortion corrections to the control waveforms.
We predistort the waveforms digitally using finite- and infinite impulse response filters applied in real time, i.e., without precompilation of the waveform, in a manner compatible with codeword-based microarchitectures~\cite{Fu17,Fu19} and feedback control.
We consistently find the reconstructed step response to be within $\sim 0.1\%$ of the ideal response in several setups and devices.

\begin{figure}
  \centering
    \includegraphics[width=\columnwidth]{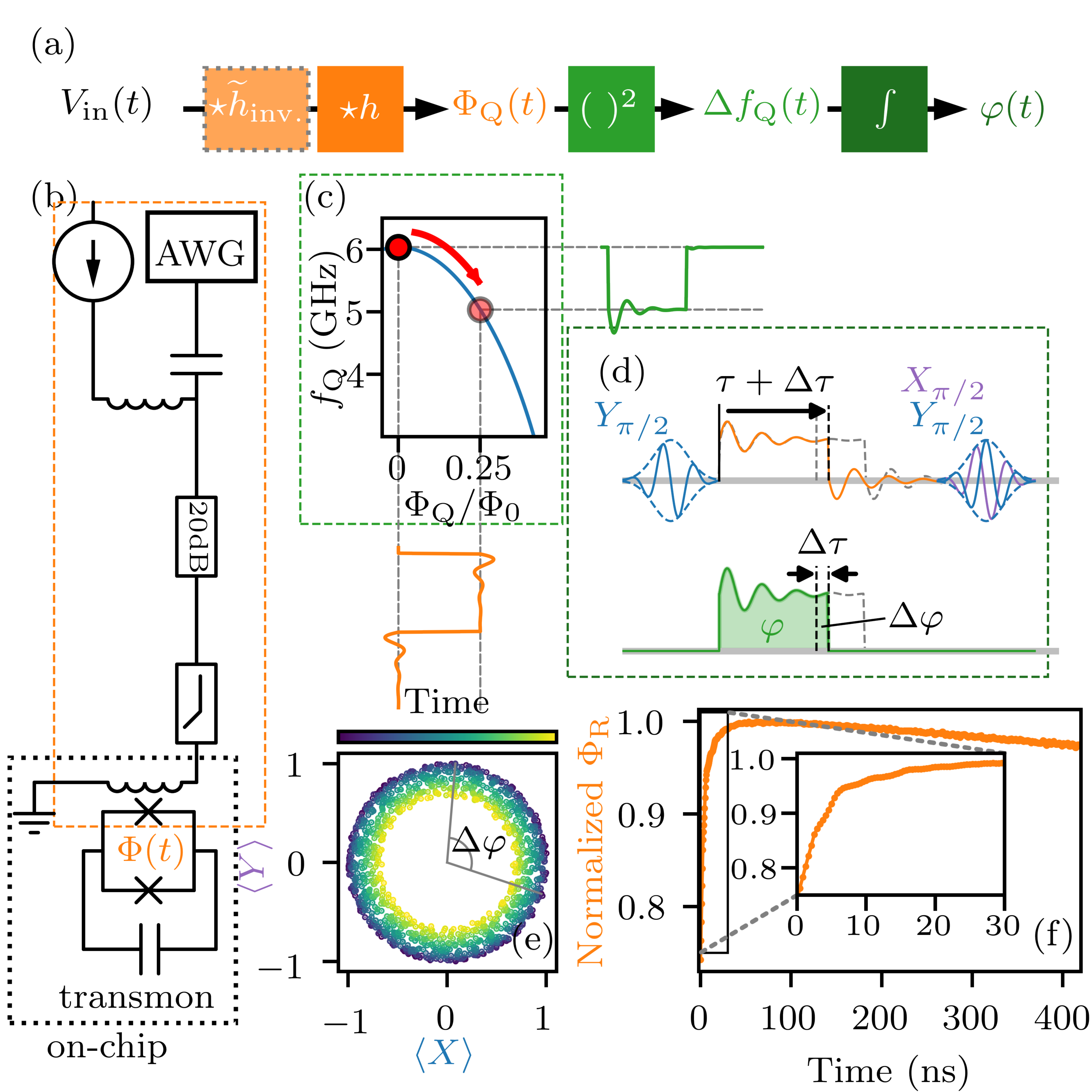}
    \caption{\label{fig:concept} Basic concept of Cryoscope.
    (a) Overview of relevant transformations involved.
    (b)    Schematic of the control line used to control the flux $\PhiQ$ through the transmon SQUID loop. A DC source and AWG combined at a bias tee at room temperature produce the static and dynamic components of $\PhiQ$.
    (c) When operating Cryoscope, the transmon is biased at its flux sweetspot and pulsed away only during the waiting interval between the $\pi/2$ pulses in a standard Ramsey-style experiment.
    (d) The difference in quantum phase $\Delta\phaseQ$ [shown in (e)] acquired by the qubit during Ramsey experiments with the flux pulse truncated after $\tTrunc$ and $\tTrunc+\Delta\tTrunc$ provides an estimate of the instantaneous qubit detuning $\detuning$ in the interval $[\tTrunc, \tTrunc+\Delta\tTrunc]$, and consequently an estimate $\PhiR$ of the instantaneous actual flux $\PhiQ$.
    The nonlinear dependence of $\detuning(\PhiQ)$ suppresses the error produced by the difference of the two turn-off transients.
    (f) Reconstructed step response of the control line, normalized to maximal flux.
  }
\end{figure}

The transition frequency $\freqQ$ of a two-junction transmon depends on the magnetic flux $\PhiQ(t)$ through its SQUID loop and for symmetric junctions is given by~\cite{Koch07}
\begin{equation}
\label{eq:flux_to_freq}
\freqQ(\PhiQ) \approx \frac{1}{h}\left(\sqrt{8\Ej \Ec \!\abs{\cos{\left(\pi \frac{\PhiQ}{\Phio}\right)}}} -\Ec\right),
\end{equation}
where $\Ec$ is the charging energy, $\Ej$ is the sum of the Josephson energies of the individual junctions, $\Phio$ is the flux quantum, and $h$ is Planck's constant.
In our system, the static and dynamic components of $\PhiQ$ are produced by a DC source and an AWG, respectively, and combined at a bias tee, all at room temperature.
Here, we use the DC source to null flux offsets, biasing the transmon at its maximal frequency, $\freqQmax\approx \frac{1}{h} \sqrt{8\Ej\Ec}-\Ec$, which functions as a sweetspot with first-order insensitivity to $\PhiQ$.
As in typical applications~\cite{Langford17,Rol19_NetZero,Sagastizabal19,Bultink19_ZZXX}, we use the AWG to flux pulse the transmon to detunings $\detuning(t)=\freqQmax-\freqQ(\PhiQ(t))$ up to $\sim 1~\GHz$, corresponding to $\sim 0.25\Phio$.

At its core, Cryoscope is a technique using Ramsey-style experiments to obtain an estimate $\PhiR(t)$ of the actual $\PhiQ(t)$ produced by an AWG pulse $\Vin(t)$.
We embed the flux pulse (with varying truncation of the input) between the two $\pi/2$ pulses, which are always separated by a fixed interval $\Tsep$.
The first $\pi/2$ pulse (around the $y$ axis of the Bloch sphere) prepares the qubit in the superposition state $(\ket{0}+\ket{1})/\sqrt{2}$.
An AWG pulse $\VinTrunc(t)$ truncated at time $\tTrunc$ produces a flux $\Phi_{\mathrm{Q}, \tTrunc} (t)$ that transforms the state to $(\ket{0}+e^{i\phaseQ_\tTrunc}\ket{1})/\sqrt{2}$, with relative quantum phase
\begin{equation}
\label{eq:cryoscope_phase_simple}
\phaseQ_\tTrunc/2\pi = \int_0^{\tTrunc}\detuning(\Phi_{\mathrm{Q}, \tTrunc} (t)) \dd{t}+\int_\tTrunc^{\Tsep}\detuning(\Phi_{\mathrm{Q}, \tTrunc} (t)) \dd{t},
\end{equation}
where we explicitly separate the contributions from the flux response up to the truncation point and the subsequent turn-off transient.
We complete the Ramsey experiment with two variants, one with the final $\pi/2$ rotation around $y$ and another with it around $x$ before measuring in order to determine the Bloch vector components $\langle X\rangle$ and $\langle Y \rangle$ from which we extract $\phaseQ_\tTrunc$.

We estimate $\PhiQ(t)$ in the small time interval $[\tTrunc, \tTrunc+ \Delta \tTrunc]$ using the following procedure.
First, we measure $\phaseQ_\tTrunc$ and $\phaseQ_{\tTrunc + \Delta \tTrunc}$ to compute
\begin{equation}
\label{eq:cryoscope_detuningEst}
\detuningEst \equiv
\frac{\phaseQ_{\tTrunc+\Delta\tTrunc}-\phaseQ_{\tTrunc}}{2\pi \Delta\tTrunc} =\frac{1}{\Delta\tTrunc} \int_\tTrunc^{\tTrunc+\Delta\tTrunc} \detuning(\Phi_{\mathrm{Q}, \tTrunc+\Delta\tTrunc}(t)) \dd{t}+\varepsilon,
\end{equation}
which gives the average detuning $\detuningAvg$ during the interval, with inaccuracy
\begin{equation}
\label{eq:cryoscope_detuningEst_varepsilon}
\varepsilon =\frac{1}{\Delta\tTrunc}\left(\int_{\tTrunc+\Delta \tTrunc}^{\Tsep} \detuning(\Phi_{\mathrm{Q}, \tTrunc+\Delta\tTrunc}(t)) \dd{t}
-  \int_\tTrunc^{\Tsep} \detuning(\Phi_{\mathrm{Q}, \tTrunc}(t)) \dd{t}\right),
\end{equation}
given by the difference in the phase contributions of the turn-off transients.

The phase contribution from the turn-off transients is suppressed because of the steep return to the first-order flux-insensitive sweetspot of the nearly quadratic $\detuning(\PhiQ)$.
Numerical simulations %~\cite{Suppmaterial} 
indicate that ${|\varepsilon|}/{\detuningEst} \lesssim 10^{-2}$--$10^{-3}$ for dynamical distortions of typically used components~\cite{Langford17,Sagastizabal19}\footnote{$\tfrac{|\varepsilon|}{\detuningEst} $ can be slightly larger for certain idealized filters such as a single-pole low-pass filter.%~\cite{Suppmaterial}
.}.
This suppression of ${|\varepsilon|}/{\detuningEst}$ would improve for higher order of nonlinearity in $\detuning(\PhiQ)$.

Finally, we obtain the reconstructed $\PhiR(t)$ by inversion of \cref{eq:flux_to_freq}.
The ability of Cryoscope to reconstruct pulses of arbitrary shape is shown in the Supplemental materials%~\cite{Suppmaterial} 
for the case of a pulse shaped as a traditional Dutch canal skyline.

We briefly discuss some technical aspects of the implementation.
We set $\Delta\tTrunc=1/2.4~\ns$, the minimum allowed by the sampling rate of the AWG (Zurich Instruments HDAWG).
The separation time $\Tsep$ is set $100~\ns$ longer than the largest chosen $\tTrunc$ to negate the need for fine timing calibrations and to be less sensitive to  residual detuning during the final rotation.
The phase $\phaseQ_\tau$ is determined by combining the $\langle X\rangle$ and $\langle Y \rangle$ data.
Before unwrapping the phase it is demodulated using the highest frequency component of a Fourier transform of the $\langle X\rangle$ and $\langle Y \rangle$ data.
A second-order Savitzky-Golay filter is then used to determine the derivative by fitting a polynomial in a small window around each data point.
The estimated detuning $\detuningEst$ is a sum of the frequency extracted using the Savitzky-Golay filter, the demodulation frequency and, when using large flux pulse amplitudes, appropriate multiples of the $1.2~\GHz$ Nyquist frequency.
The Nyquist order can be determined by acquiring Cryoscope traces for square pulses with different amplitudes and observing when the mean frequency wraps as the pulse amplitude is increased.
Because distortions can cause the instantaneous detuning to be slightly lower or higher than the mean detuning, amplitudes close to the Nyquist wrapping should be avoided.

As a first demonstration of Cryoscope, we measure the voltage-to-flux step response $\step$ of the control line.
The result shown in \Cref{fig:concept}(f) reveals clear deviations from the ideal, with dynamics on timescales comparable to typical pulse durations ($\sim 40~\ns$) and much longer.
These dynamics are the result of compounded linear dynamical distortions and thus can be described by convolution of the input $\Vin(t)=V_0 \cdot \heaviside$ (where $\heaviside$ is the Heaviside step function) with the system impulse response $\imp$, $\PhiQ (t)=\imp \star \Vin (t)$.
We furthermore assert that the system is causal so that $\step = 0 $ for $t<0$.

As an application of Cryoscope, we make iterative use of real-time digital filtering (available in the AWG) and Cryoscope to improve the step response.
The goal of this procedure is to determine the filter $\impFilt =\impPseudoInv$ that best inverts $\imp$ such that the corrected step response $\stepCorr = \impFilt \star \step$ approximates $\heaviside$ as close as possible.

First, several first-order infinite impulse response (IIR) filters are applied to eliminate dynamics on timescales longer than $30~\ns$.
The IIR filters are designed to each correct a step response of the form $\step = g(1+Ae^{-t/\tIIR})\cdot\heaviside$, where $A$ is the amplitude coefficient, $\tIIR$ is the time constant of the filter and $g$ is a gain constant.
The coefficients of the filters are determined by performing a least-squares optimization of a prediction of $\stepCorr$ based on a model of the IIR filters and the measured $\step$.
Because the IIR filters are applied in real-time on the hardware, there are small differences between the ideal filter and the implementation which are taken into account in the model.%~\cite{Suppmaterial}.
We typically require 3--5 such IIR filters in order to correct $\step$ between $30-200~\ns$
Cryoscope is used to evaluate the corrections of the IIR filters~[\cref{fig:dist_corr}(a)] and shows a reconstruction in which the slow dynamics are corrected to within $\sim0.1\%$.

Next, a finite impulse response (FIR) filter is used to correct for the remaining short ($<30~\ns$) timescale dynamics.
The FIR filter is described by 40 parameters that in turn describe the 72 coefficients ($30~\ns$) of the filter.%~\cite{Suppmaterial}.
The values are found by minimizing the distance between the predicted signal and the ideal step response using the CMA-ES algorithm~\cite{Hansen09}.
A third Cryoscope measurement is performed to test the accuracy of the corrections.
This final step can be used to iteratively fine tune the FIR coefficients if required.
No such iterations were required to achieve a reconstructed step response accurate to $\sim0.1\%$ shown in~\cref{fig:dist_corr}(a).

To independently characterize the corrections, we perform a chevron experiment without and with the predistortions applied [\cref{fig:dist_corr}(b,c)].
In this experiment, two qubits ($q_1$ and $q_0$) are prepared in the $\ket{11}$ state using $\pi$ pulses, a square flux pulse of varying duration and amplitude is applied to the higher frequency qubit ($q_0$) to tune $\ket{11}$ into (near) resonance with $\ket{02}$, the same interaction that is exploited to realize a CZ gate.
With no predistortions applied [\cref{fig:dist_corr}(b)], the pattern of $q_1$ population as a function of pulse amplitude (horizontal axis) and duration (vertical axis) is visibly asymmetric -- fringes on the right-hand side are more visible, and the pattern bends towards large pulse amplitudes for short pulse durations.
These two features are signatures of the finite rise time of the applied pulse.
In contrast, when predistortions are applied [\cref{fig:dist_corr}(c)], the pattern is almost perfectly left-right symmetric, both in terms of visibility and shape, indicating a near-perfect rectangular pulse.
Using Cryoscope, we can predict the pulse amplitude that results in exact $\ket{11}$--$\ket{02}$ degeneracy at every point in time.
The prediction [white curve in \cref{fig:dist_corr}(b, c)] overlaps with the path along which the oscillations are slowest, providing an independent verification (although less quantitative) of our method.

\begin{figure}
  \centering
    \includegraphics[width=\columnwidth]{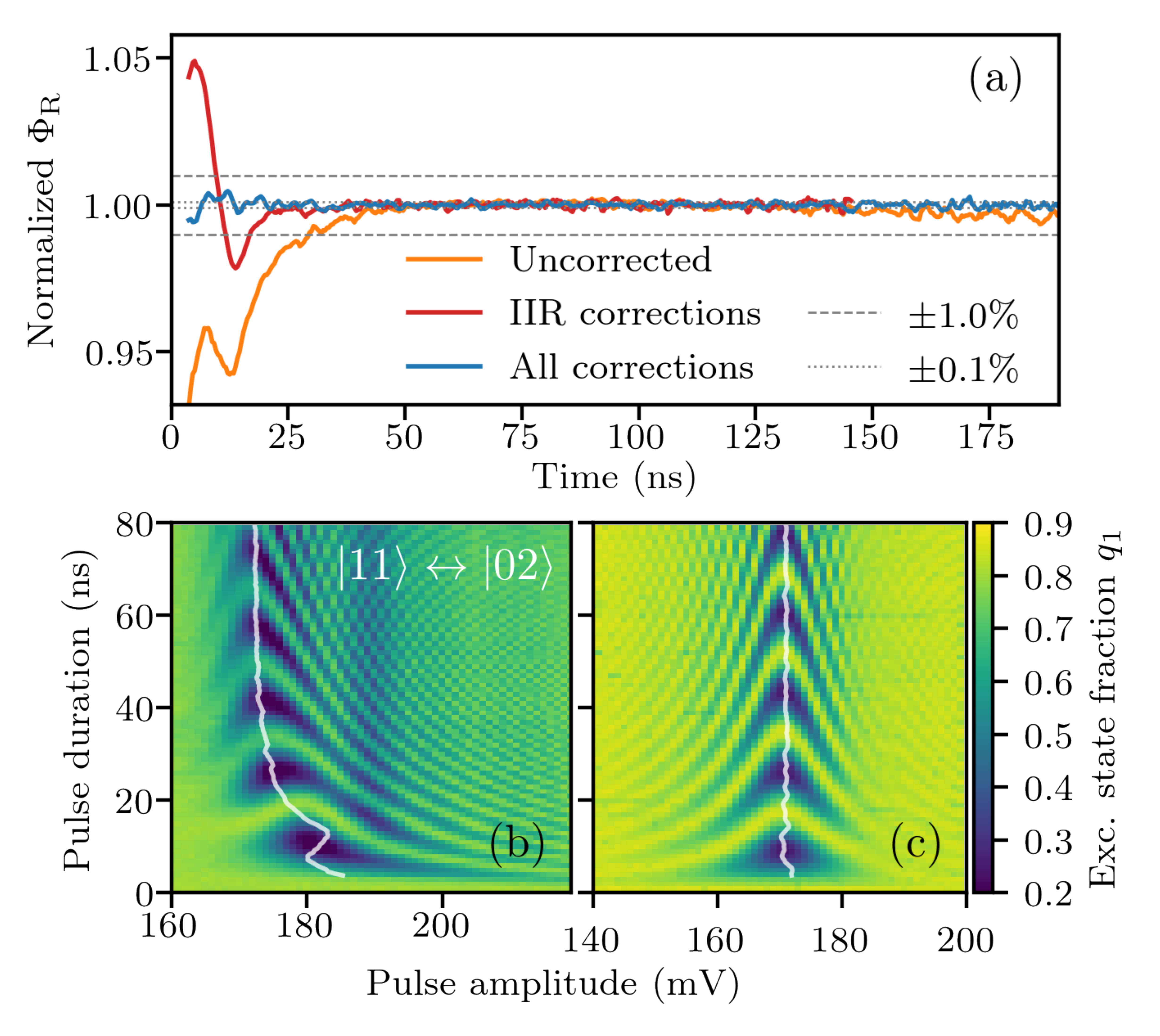}
  \caption{\label{fig:dist_corr} Reconstructed step response without and with distortion corrections (for a qubit on a different device from that of \cref{fig:concept}) normalized to flux between 40 and $125~\ns$.
  (a) Cryoscope measurements of uncorrected (orange) and corrected step responses with IIR corrections only (red) and FIR and IIR corrections (blue).
  (b-c) Chevron experiments without and with predistortion corrections (not corrected for readout error).
  The overlaid curve indicates resonance between $\ket{11}$ and $\ket{02}$, predicted using the step response reconstructed with Cryoscope.
  See text for details.
  }
\end{figure}

Having established the ability to measure and correct distortions, we investigate the sensitivity of Cryoscope.
\Cref{fig:SNR}(a) presents the unprocessed measurement of $\langle X \rangle$ for three values of qubit detuning during the rectangular pulse.
In all cases we observe decaying oscillations. The decay is faster the larger the pulse amplitude due to reduced coherence of the qubit further away from sweetspot.
The reconstructed instantaneous flux in a $100-200~\ns$ window [\cref{fig:SNR}(b,c)] fluctuates around the mean value, in a range decreasing with the amplitude of the rectangular pulse.
We interpret that for larger detuning the qubit precession is faster, resulting in a larger phase acquired between subsequent time steps and consequently yielding a more accurate measurement of the instantaneous detuning relative to nearly the same sampling noise.

We define a signal-to-noise ratio to quantify the influence of dephasing and precession rate on Cryoscope sensitivity,
\begin{equation}
\label{eq:SNR}
\mathrm{SNR} = \frac{\PhiRmn}{\PhiRstd}.
\end{equation}
We define as signal the mean amplitude of the optimally corrected, reconstructed flux $\PhiRmn$ and as noise the standard deviation $\PhiRstd$
The SNR is experimentally determined for several time windows and amplitudes of the rectangular flux pulse~[\cref{fig:SNR}(d)].
We perform 10 Cryoscope experiments for every data point to extract $\PhiRmn$ and $\PhiRstd$ in the relevant time interval.
In the $100-200~\ns$ window, SNR increases quadratically with pulse amplitude, indicating that detuning increases, while the qubit coherence is not affected on this short timescale.
In contrast, the increase of SNR is slower for the other time windows. In particular, for the $1200-1300~\ns$ window, the SNR reaches a maximum for pulse amplitude $\PhiQ \approx 0.17~\Phio$.
The maximum indicates the configuration in which the benefit of increased precession rate balances out the drawback of the reduced qubit coherence (due to increased sensitivity to flux noise).

The SNR is also affected by acquisition and processing parameters.
The former is the precision with which the qubit state can be determined, which depends on the number of averages and the readout fidelity.
The latter is a matter of applied data filtering and can be adjusted depending on the temporal resolution demanded.

All these factors can be combined in a model yielding%~\cite{Suppmaterial} yielding
\begin{equation}
\SNR = \aSNR \PhiQ^2 \exp \left( -(\Gamma_0 + 2 a \Gamma_1 \PhiQ) t \right),
\end{equation}
where $t$ is the time of reconstruction, $\aSNR$ accounts for sampling noise and filtering effects in data processing, $\Gamma_0$ is a sweetspot dephasing rate, $\Gamma_1$ quantifies the power of flux noise and the qubit detuning from sweetspot is $\detuning (\PhiQ) = a \PhiQ^2$.
The interplay between quadratic and exponential terms in $\PhiQ$ represents the trade-off between improved sensitivity to the shape of flux pulse versus reduced signal visibility due to dephasing.
The prefactor $\aSNR$ can be increased by averaging more or alternatively improving the readout fidelity.
We fix values of $a$ and $\Gamma_0$ based on independent measurements%~\cite{Suppmaterial} 
and perform a fit of the two-parameter model ($\aSNR$ and $\Gamma_1$), finding a good agreement with the data [\cref{fig:SNR}(d)].

\begin{figure}
  \centering
    \includegraphics[width=\columnwidth]{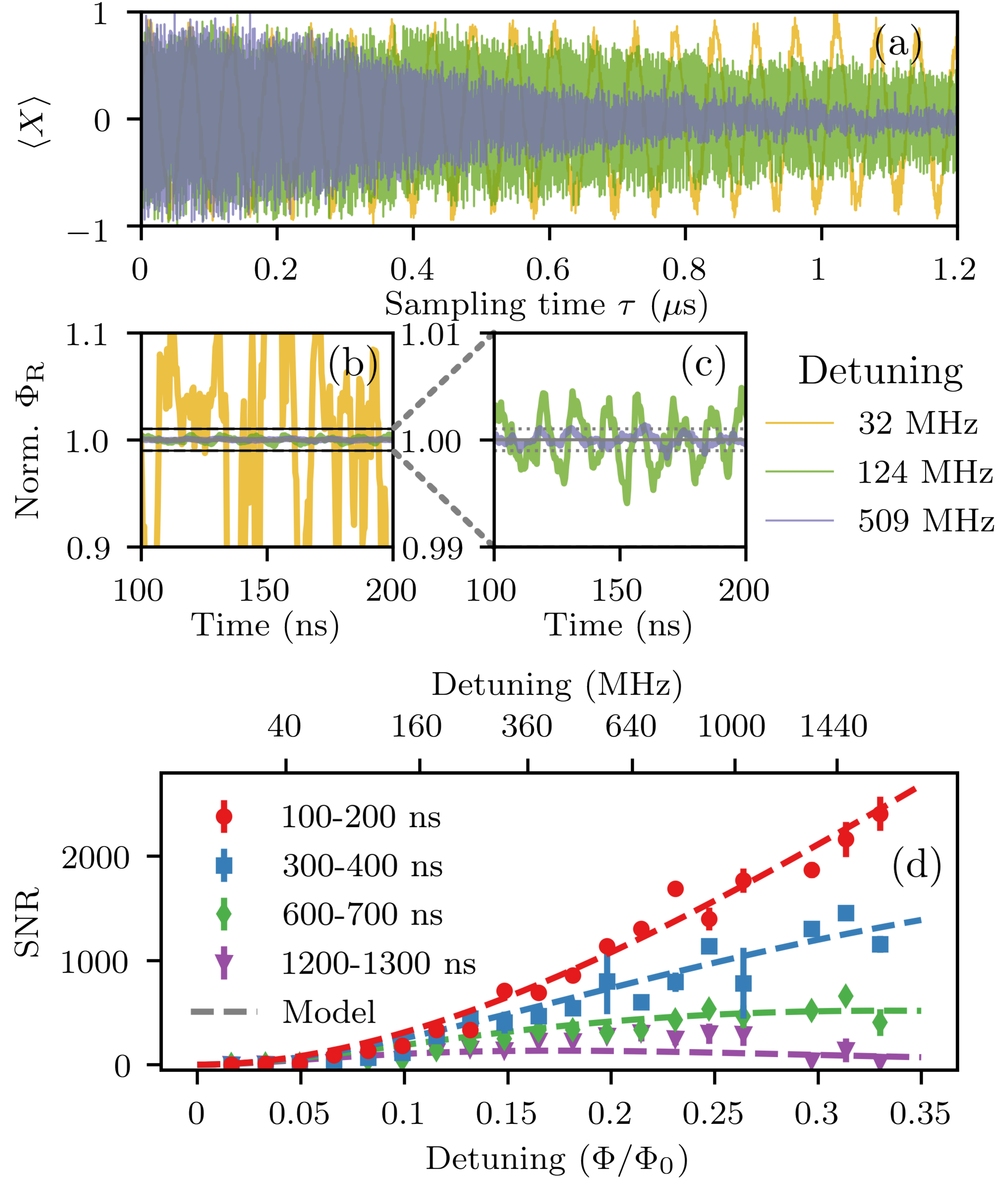}
  \caption{\label{fig:SNR} Cryoscope signal-to-noise ratio. (a) Raw measurements of $\langle X \rangle$ for individual Cryoscope traces using different detuning. (b, c) Zoom of reconstructed signal (normalized to the mean flux). The dotted curves denote deviations of 0.1\%. (d) SNR at various timescales and detunings.
  }
\end{figure}

In conclusion, we have demonstrated a method capable of sampling on-chip flux pulses by exploiting the nonlinear flux dependence of transmon frequency.
This characterization method is straightforward to use and generalizable to any qubit system with baseband control of the qubit frequency and a sweetspot with respect to the control variable.
Furthermore, we have demonstrated the capability to correct distortions as demonstrated by a reconstructed step response accurate to $\sim0.1\%$.
The identified corrections were applied in real time, making the correction method compatible with an instruction-based control architecture~\cite{Fu17,Fu19}.
Cryoscope has already been used to tune-up fast, high-fidelity, and low-leakage CZ gates for a QEC experiment~\cite{Rol19_NetZero, Bultink19_ZZXX} and parametrized iSWAP interactions in a variational quantum eigensolver~\cite{Sagastizabal19,OBrien19}.

\section*{Supplemental materials}
The supplemental material provides experimental details and derivations supporting claims made in the main text. 
First, we describe the experimental setup.
We next discuss technical details of Cryoscope.
The third section details a simple model for the signal-to-noise ratio of Cryoscope.
Next, we provide details on the hardware implementations of the FIR and IIR filters used to correct distortions in real time.
Finally, we provide experimental data demonstrating the ability to use Cryoscope to reconstruct an arbitrary signal.

\begin{acknowledgments}
This research is supported by the Office of the Director of National Intelligence (ODNI), Intelligence Advanced Research Projects Activity (IARPA),
via the U.S. Army Research Office grant W911NF-16-1-0071, by Intel Corporation, and by the ERC Synergy Grant QC-lab.
The views and conclusions contained herein are those of the authors and should not be interpreted as necessarily representing the official policies or endorsements, either expressed or implied, of the ODNI, IARPA, or the U.S. Government.
\end{acknowledgments}

% \bibliographystyle{apsrev4-1}
% \bibliography{../../../../Paper_resources/References/References_cQED}
%merlin.mbs apsrev4-1.bst 2010-07-25 4.21a (PWD, AO, DPC) hacked
%Control: key (0)
%Control: author (72) initials jnrlst
%Control: editor formatted (1) identically to author
%Control: production of article title (-1) disabled
%Control: page (0) single
%Control: year (1) truncated
%Control: production of eprint (0) enabled
%

\clearpage

\renewcommand{\theequation}{S\arabic{equation}}
\renewcommand{\thefigure}{S\arabic{figure}}
\renewcommand{\thetable}{S\arabic{table}}
\renewcommand{\bibnumfmt}[1]{[S#1]}
\renewcommand{\citenumfont}[1]{S#1}
\setcounter{figure}{0}
\setcounter{equation}{0}
\setcounter{table}{0}

\onecolumngrid
\section*{Supplemental material for ``Time-domain characterization and correction of on-chip distortion of control pulses in a quantum processor''}

This supplement provides experimental details and derivations supporting claims made in the main text.
First, we describe the experimental setup.
We then discuss the limitations of the Cryoscope, showing how undesired distortions are suppressed for a typical step response and how the  nonlinear response of the qubit to flux helps in reconstructing the step response.
The third section details a simple model that describes the signal-to-noise ratio of the experiment.
Next, we provide details on the hardware implementations of the FIR and IIR filters used to correct distortions in real time.
Finally, we provide experimental data demonstrating the ability to use Cryoscope to reconstruct an arbitrary signal.

\section{Device and experimental setup}
The data shown in this letter were acquired using two devices mounted in different dilution refrigerators.
In all experiments, a Zurich Instruments HDAWG equipped with real-time digital filters was used to generate the flux pulses.
The output of the AWG was connected to the RF port of a Mini-Circuits ZFBT 6GW+ bias tee while the DC port was connected to a DC current source.
The RF+DC port of the bias tee was connected to the flux control line entering the fridge.
The flux control line contains a $20~\dB$ attenuator at the $4~\K$ stage as well as a Mini-Circuits VLFX1050 low-pass filter and a homebuilt eccosorb filter before being connected to the flux control line on the device.
The control-line coaxial cables between $4~\K$ and mixing chamber plate were superconducting (NbTi, inner and outer conductor) for one device and stainless steel (inner and outer) for the other.
In all cases it was possible to correct distortions to within $\sim 0.1\%$.

\section{Limitations of the Cryoscope}
In this section we first investigate the accuracy of the  Cryoscope for a physically motivated step response including distortions due to AWG bandwidth,  bias tee, skin effect, and on-chip response.
We show that the inaccuracy is small using typical distortion parameters.
Next, we investigate the effects of a single-pole low-pass filter for which the error in the reconstruction is significant on the timescale of the filter.

Our analysis is based on a numerical calculation of the acquired relative phase $\phaseQ_{\tTrunc}$, yielding a noiseless Cryoscope measurement.
Specifically,
\begin{equation}
    \label{eq:nominal_cryoscope_phase}
    \phaseQ_{\tTrunc} = 2\pi\int\limits_0^\infty a \left[ \Phi_s \left( \step - s(t-\tTrunc) \right) \right] ^k dt,
\end{equation}
where $\Phi_s$ is the amplitude of the applied square flux pulse, $\step$ is the step response and $a$ parametrizes the dependence between magnetic flux $\Phi$ and qubit detuning in the quadratic approximation: $\detuning = a\Phi^k$ with $k=2$.
Because the phases are calculated in simulation it is  possible to extract the contributions to $\detuningEst$:

\begin{equation}
\label{eq:cryoscope_detuning_contrib}
\detuningEst (\tTrunc, \Delta\tTrunc) \equiv
\frac{\phaseQ_{\tTrunc+\Delta\tTrunc}-\phaseQ_{\tTrunc}}{2\pi \Delta\tTrunc} =
    \underbrace{\frac{1}{\Delta\tTrunc}\int_\tTrunc^{\tTrunc+\Delta\tTrunc} \detuning(\Phi_{\mathrm{Q}, \tTrunc+\Delta\tTrunc}(t)) \dd{t}}_{\detuningAvg} +
\overbrace{
    \underbrace{\frac{1}{\Delta\tTrunc}\int_{\tTrunc+\Delta \tTrunc}^{\Tsep} \detuning(\Phi_{\mathrm{Q}, \tTrunc+\Delta\tTrunc}(t)) \dd{t}}_{\varepsilon_{\tTrunc+\Delta\tTrunc, \mathrm{Off}}}
-   \underbrace{\frac{1}{\Delta\tTrunc}\int_\tTrunc^{\Tsep} \detuning(\Phi_{\mathrm{Q}, \tTrunc}(t)) \dd{t}}_{\varepsilon_{\tTrunc, \mathrm{Off}}}}^{\varepsilon}.
\end{equation}

\subsection{Cryoscope reconstruction of a typical step response}

An overview of the distortion models used can be found in \cref{tab:physical_dist_model}.
The response of the HDAWG is taken into account by performing a convolution with an impulse response extracted from the measured step response.
This step response was measured when the HDAWG was operated in amplified mode and shown in \cref{fig:HDAWG_step}.
The step response of the bias tee is modeled as a single exponential high-pass filter of the form $\step = e^{-t/\tHP} \cdot\heaviside$ and several exponential filters of the form $\step = 1+ Ae^{-t/\tIIR}\cdot\heaviside$, where $\tIIR$ and $\tHP$ are the relevant time constants, $A$ is an amplitude coefficient and $\heaviside$ is the Heaviside step function.
The coefficients used are based on a measured step response for a bias tee and are the same as in Ref.~\onlinecite{Sagastizabal19_S}.
We note that the coefficients are known to vary slightly between different bias tees of the same model.
The skin effect is modeled according to Ref.~\onlinecite{Wigington57} with an attenuation of $\alpha_\GHz=2.1~\dB$ at $1~\GHz$.
The signal is filtered with a Savitsky-Golay filter in order to determine the derivative and the impulse response.

\begin{figure}
  \centering
    \includegraphics[width=0.45\textwidth]{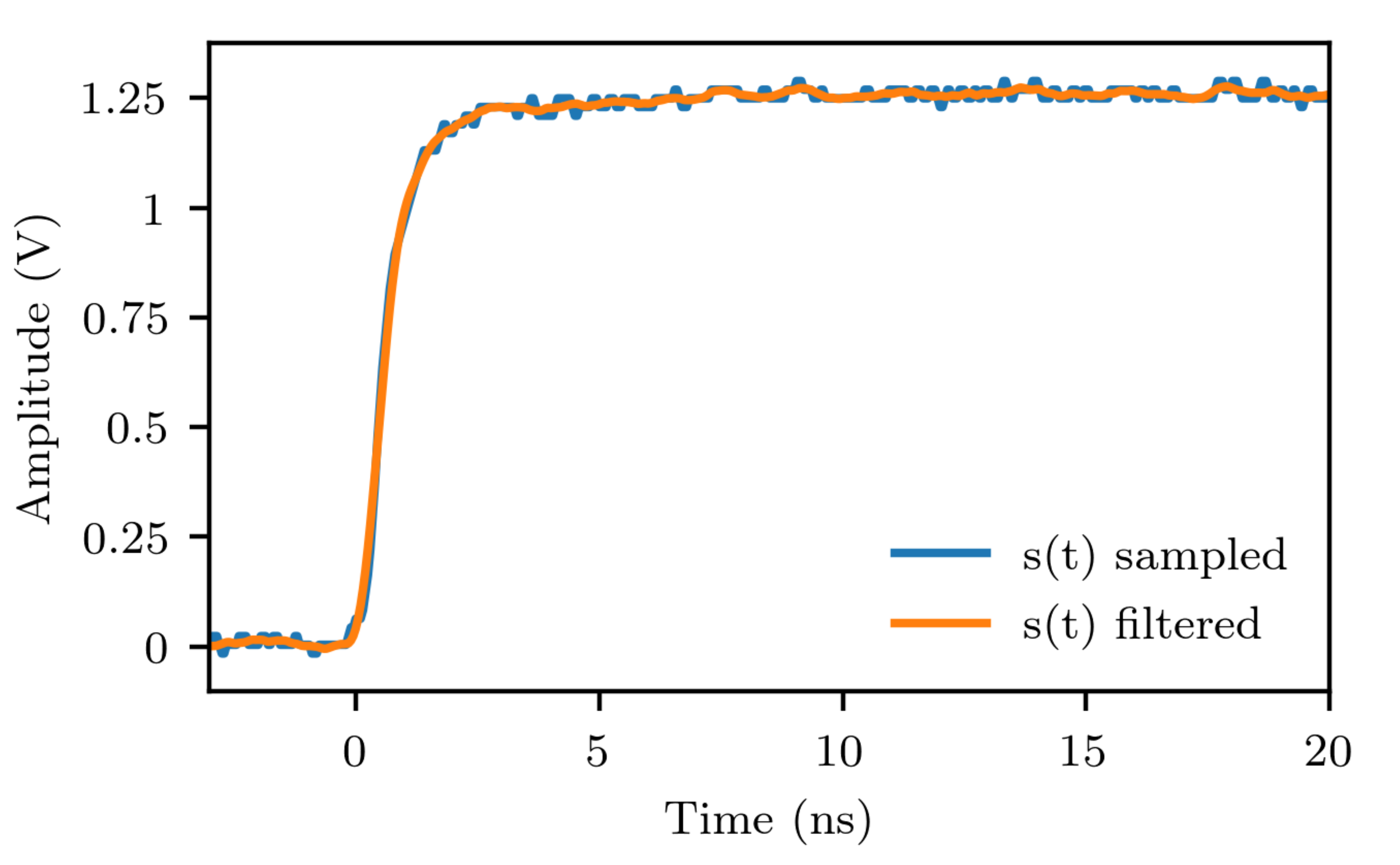}
    \caption{\label{fig:HDAWG_step}Measured step response of the HDAWG in amplified mode, measured using a Rohde \& Schwarz RTO1024 oscilloscope (blue).
    The signal is filtered with a Savitsky-Golay filter (orange) in order to determine the impulse response from the derivative.}
\end{figure}

\begin{table}[htb]
\begin{tabular}{|c|c|c|c|}
\hline
\textbf{Effect}  & \textbf{Model}& \textbf{\begin{tabular}[c]{@{}l@{}}Model\\ parameters\end{tabular}} & \textbf{Notes}\\ \hline
AWG response     & $\star h_\mathrm{AWG}$& -  & \makecell{Measured step response of \\HDAWG in amplified mode (\cref{fig:HDAWG_step})}\\ \hline
Bias tee         & $\step=(e^{-t/\tHP})\cdot\heaviside$ & $\tHP=41~\us$ & Mini-Circuits ZFBT-6GW+\\ \hline
Bias tee         & $\step=(1+A \cdot e^{-t/\tIIR})\cdot\heaviside$ & \makecell{$\tIIR = 15~\us$\\ $A=0.13 $} & Ref.~\onlinecite{Sagastizabal19_S}\\ \hline
Bias tee         & $\step=(1+A \cdot e^{-t/\tIIR})\cdot\heaviside$ & \makecell{$\tIIR = 6.4~\us$\\ $A=0.99 $} & Ref.~\onlinecite{Sagastizabal19_S} \\ \hline
Skin effect      & $\step=(1-\mathrm{erfc} \left(\alpha_{\mathrm{GHz}}/21\sqrt{t}\right))\cdot\heaviside$ & $\alpha_{\mathrm{GHz}}=2.1\dB$& \makecell{Model according to Ref.~\onlinecite{Wigington57}}
\\ \hline
On-chip response & $\step =( 1+A\cdot e^{-t/\tIIR})\cdot\heaviside$& \makecell{$\tIIR=2~\ns$\\ $A=0.6$}&  \\ \hline
\end{tabular}
\caption{Overview of the distortion models used and their coefficients. \cref{fig:physical_model_correspondence} illustrates the cumulative influence of all listed effects and their reconstruction using Cryoscope. \label{tab:physical_dist_model}}
\end{table}

\begin{figure}
  \centering
    \includegraphics[width=1\textwidth]{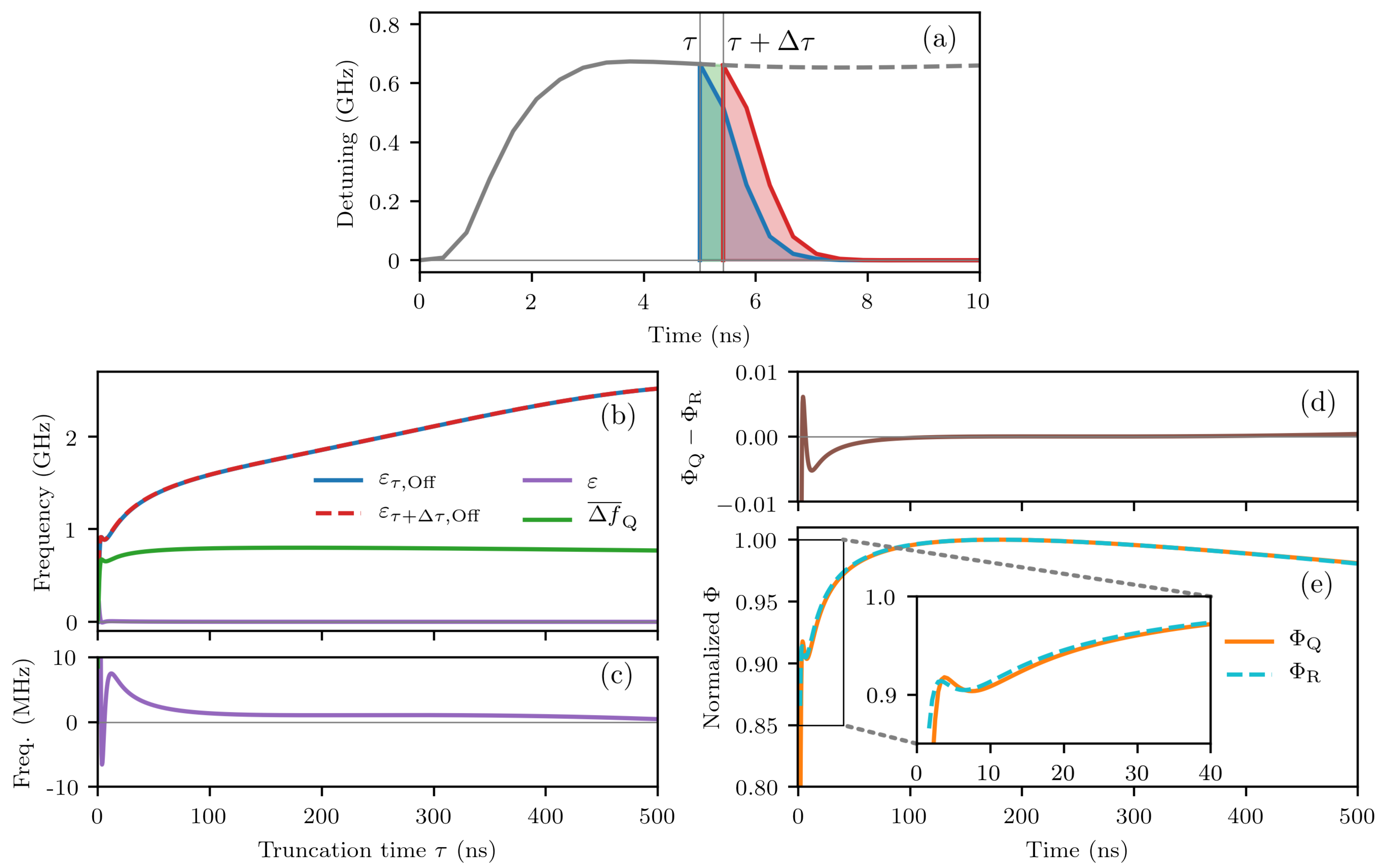}
  \caption{\label{fig:physical_model_correspondence} Simulated reconstruction of a typical step response (cumulative effect of models of \cref{tab:physical_dist_model}) using Cryoscope.
  (a) Detuning of the qubit when applying square pulses truncated at $\tTrunc$ and $\tTrunc+\Delta \tTrunc$.
  Shaded areas illustrate the contributions to $\detuningEst$ from $\detuningAvg$ (green), $\varepsilon_{\tTrunc, \mathrm{Off}}$ (blue) and $\varepsilon_{\tTrunc+\Delta\tTrunc, \mathrm{Off}}$ (red).
  (b, c) Contributions to $\detuningEst$ from the true detuning (green), the individual turn-off transients (blue and red), and the difference of turn-off transients (purple).
  (e, f) Comparison of the reconstructed flux $\PhiR(t)$ to the true flux $\PhiQ(t)$.
  }
\end{figure}

We model the effect of distortions on square pulses (truncated at time $\tTrunc$ and $\tTrunc +\Delta\tTrunc$) that detune the qubit by $\detuning = 800~\MHz$~\cref{fig:physical_model_correspondence}.
The contribution of the individual turn-off transients $\varepsilon_{\tTrunc, \mathrm{Off}}$ and $\varepsilon_{\tTrunc+\Delta \tTrunc, \mathrm{Off}}$ is typically significantly larger than $\detuningAvg$ as it takes multiple $\Delta \tTrunc$ for the qubit to return to the sweetspot.
However, their difference $\varepsilon$ is smaller; with the exception of the first few samples, $\epsilon \lesssim 8~\MHz\approx\detuningAvg\cdot 10^{-2}$~[\cref{fig:physical_model_correspondence}(b, c)].
We next use the acquired phases as input for the Cryoscope analysis to obtain the reconstructed flux $\PhiR(t)$ and compare it to the true flux $\PhiQ(t)$~[\cref{fig:physical_model_correspondence}(d, e)].
We observe a matching of $\PhiR(t)$ to $\PhiQ(t)$ better than $1\%$ for $t>3~\ns$.
Note that the data shown in \cref{fig:physical_model_correspondence}(d, e) is normalized to the maximal flux.

\subsection{Cryoscope reconstruction of a single-pole low-pass filter step response}
For completeness, we demonstrate that Cryoscope may reconstruct the step response poorly for specific filters.
A simple example is the single-pole low-pass filter, whose step response is
\begin{equation}
    \label{eq:LP}
    s_\mathrm{LP}(t) = (1-e^{-t/\tau_{\mathrm{LP}}}) \cdot\heaviside,
\end{equation}
where $\tau_{\mathrm{LP}}$ is the time constant.
Such a filter does not accurately represent our setup but is easy to describe analytically and therefore is a good choice to demonstrate the origin of potential errors and to show the relevance of the nonlinear qubit response to flux.
We find that the reconstructed step response $s_{R,\mathrm{LP}}(t)$ differs from $s_\mathrm{LP}(t)$ by more than 1\% for $t \lesssim 4\tau_{\mathrm{LP}}$.

We also use this simple example to show that Cryoscope is more accurate for higher degrees of nonlinearity.
Specifically, we calculate $s_{R,\mathrm{LP}}(t)$ for different forms of qubit detuning on flux: $\Delta f(\Phi) = a \Phi^k$ where $k  \in \mathbb{Z}^+$.

In general, the phase $\phaseQ_\tTrunc$ (setting $\Tsep=\infty$) expressed in terms of the impulse response $h = \dd{s}/\dd{t}$ is
\begin{equation}
    \phaseQ_\tau =
    2\pi a \int\limits_0^\infty \left[\int\limits_0^\infty h(t-t') \dd t' - \int\limits_0^\infty h(t-\tau-t') \dd t' \right]^k \dd t =
    2\pi a \int\limits_0^\tau \left[\int\limits_0^t h(t-t') dt' \right]^k \dd t +
    2\pi a \int\limits_\tau^\infty \left[\int\limits_0^\tau h(t-t') dt' \right]^k \dd t,
\end{equation}
 while the reconstructed step response is given by
\begin{equation}
    s_R (\tau) = \left( \frac{2\pi}{a} \frac{\dd \phaseQ}{\dd \tau} \right)^{1/k} =
    \left( \int\limits_\tau^\infty \left\lbrace h(t-\tau) \left[ \int\limits_0^t h(t-t') \dd t' \right]^{k-1} \right\rbrace \dd t \right)^{1/k}.
\end{equation}
For the single-pole low-pass filter,
\begin{equation}
    \label{eq:RC_measured}
    s_{R,\mathrm{LP}} (t) =
    \left[1-  e^{-t/\tau_\mathrm{LP}} \right]^{\frac{k-1}{k}}\heaviside  =
    \left\lbrace \begin{array}{lll}
     (1-  e^{-t/\tau_\mathrm{LP}})\cdot\heaviside & \mathrm{for} & k\rightarrow \infty \\
     \heaviside & \mathrm{for} & k=1
     \end{array}
     \right.
\end{equation}
We observe that, $s_{R,\mathrm{LP}}$ matches $s_\mathrm{LP}$ in the limit $k \rightarrow \infty$.
On the other hand, for $k=1$ (i.e. linear dependence of qubit frequency on flux) the reconstruction gives $\heaviside$, completely missing the dynamics.

By directly comparing \cref{eq:LP} to \cref{eq:RC_measured} we can place an upper bound on the Cryoscope inaccuracy:
\begin{equation}
    0 \leq s_{R,\mathrm{LP}}(t) - s_\mathrm{LP}(t) \leq \frac{1}{k}e^{-t/\tau_{\mathrm{LP}}}.
\end{equation}
The factor $1/k$ clearly shows that the nonlinear dependence on flux is essential for Cryoscope to work and that a higher power dependence increases its accuracy.

The reduced inaccuracy with higher order $k$ can intuitively be understood by considering that $\varepsilon$ is determined by contributions during the turn-off transients (see \cref{eq:cryoscope_detuning_contrib}).
In the case of a linear dependence, i.e., $k=1$, contributions during the turn-off transients are weighted as strong as contributions during the pulse, making it impossible to reconstruct the waveforms.
With a parabolic dependence, i.e., $k=2$, the contributions during the off-transients are suppressed.
The higher the order $k$, the more strongly $\varepsilon$ is suppressed.

\section{Cryoscope signal-to-noise ratio}

Taking the expectation values  $\langle X \rangle$ and $\langle Y \rangle$ in Cryoscope, denoted $x$ and $y$ for brevity,
the probability density $p(x,y)$ of measuring $(x,y)$, given that the true values are $(x_0,y_0)$, is assumed to be normally distributed with standard deviation $\sigma$ dependent on readout fidelity and number of averages,
\begin{equation}
    p(x,y) = \frac{1}{2 \pi \sigma^2}\exp\left( -\frac{(x-x_0)^2 + (y-y_0)^2}{2 \sigma^2} \right).
    \label{pxy}
\end{equation}

\begin{figure}[tbh]
    \label{fig:SNR_variables}
    \centering
    \includegraphics[width=0.35\textwidth]{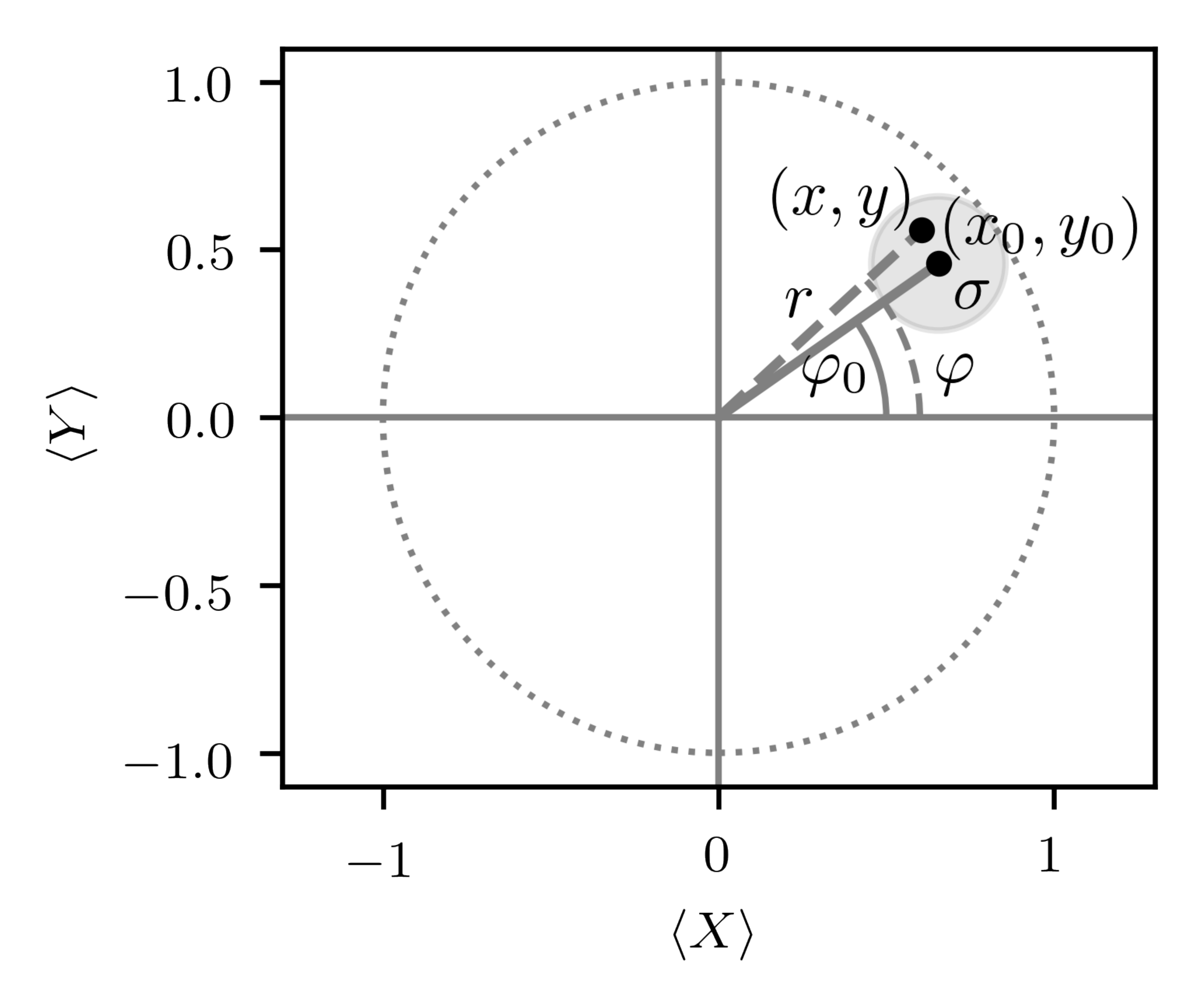}
    \caption{
    Illustration clarifying the variables used.
    The probability of measuring $(x,y)$, given that the actual values are $(x_0,y_0)$, is taken to be normally distributed with standard deviation $\sigma$ denoted by the grey circle. }
\end{figure}

The values $x$ and $y$ can be represented as an amplitude $r$ and a phase $\phaseQ$.
Assuming a perfect (non-distorted) square pulse, the value $r$ is affected by the dephasing of the qubit $r_{\tTrunc} = \exp\left(-(\tTrunc/\Ttwostar(\PhiQ))^\alpha\right)$ where $\Ttwostar(\Phi)$ is the flux dependent dephasing rate and $\alpha \in [1,2]$.
Meanwhile, the phase $\phaseQ_\tTrunc=2 \pi \tTrunc \detuning(\PhiQ)$, where $\detuning(\PhiQ)$ is the qubit detuning from the operating point as a function of flux-pulse amplitude.
The phase is measured with some error $\sigma_{\phaseQ}$ that depends on $r$ and $\sigma$.
Ultimately, to reconstruct $\PhiR$, we are interested in the phase change between two measurements with slightly different $\tTrunc$, $\phaseDiff =  \phaseQ_{\tTrunc+\Delta\tTrunc}-\phaseQ_{\tTrunc}$, with error $\sigma_{\phaseDiff}$.

\subsection{Error propagation}

Without loss of generality, $\phaseQ_0$ can be set to 0, setting $y_0 = 0$ in Eq.~\eqref{pxy}.
Rewriting $p(x,y)$ in spherical coordinates,
\begin{equation}
    p(r,\phaseQ) = \frac{1}{2 \pi \sigma^2}\exp\left( -\frac{r^2 - 2 r x_0 \cos(\phaseQ) + x_0^2}{2 \sigma^2} \right),
\end{equation}
where the subscript from $r_{\tTrunc}$ and $\phaseQ_{\tTrunc}$ is dropped for brevity.
As we are only interested in the error of the phase, we can integrate to find
\begin{equation}
\begin{split}
    p(\phaseQ) = \int\limits_0^\infty p(r,\phaseQ) r dr &=
    \frac{1}{2 \pi} \exp \left( -\frac{x_0^2}{2 \sigma^2}    \right) \\
    &+\frac{\cos(\phaseQ)}{\sqrt{8 \pi}} \frac{x_0}{\sigma}
     \exp \left( -\frac{x_0^2 \sin^2(\phaseQ)}{2 \sigma^2} \right)
     \left[ \erf\left( \frac{x_0 \cos(\phaseQ)}{\sqrt{2} \sigma}\right) +1 \right].
     \label{eq:pphi}
\end{split}
\end{equation}

When the the visibility of the Ramsey oscillations $x_0$  is much larger than $\sigma$ this simplifies to
\begin{equation}
\begin{split}
    p(\phaseQ) = \frac{\cos(\phaseQ)}{\sqrt{8 \pi}} \frac{x_0}{\sigma}
     \exp \left( -\frac{x_0^2 \sin^2(\phaseQ)}{2 \sigma^2} \right)
     \left[ \erf\left( \frac{x_0 \cos(\phaseQ)}{\sqrt{2} \sigma}\right) +1 \right].
\end{split}
\end{equation}

Because we have set $\phaseQ_0=0$ and $x_0\ll\sigma$, the small-angle approximation ($\phaseQ\ll1$) can be made, so that this simplifies further to
\begin{equation}
    p(\phaseQ) = \frac{x_0}{\sqrt{2 \pi} \sigma} \exp\left( -\frac{x_0^2 \phaseQ^2}{2 \sigma^2} \right).
     \label{pphi_small}
\end{equation}
Since the distribution is normal, we conclude that
\begin{equation}
    \sigma_\phaseQ =  \frac{\sigma}{x_0}\quad \mathrm{and} \quad    \sigma_{\phaseDiff} = \frac{\sqrt{2} \sigma}{x_0}.
\end{equation}

\subsection{SNR formula}
Ultimately, the SNR of Cryoscope is affected by the following factors:
\begin{enumerate}
    \itemsep=2pt
    \item Readout fidelity and averaging, captured by $\sigma$;
    \item Flux dependent qubit dephasing $\Ttwostar(\PhiQ)$, affecting  visibility
        $r = x_0 = \exp\left(-(\tTrunc/\Ttwostar(\PhiQ))^\alpha\right)$, with $1 \leq \alpha \leq 2$;
    \item Rate at which the phase is acquired, proportional to $\detuning(\PhiQ)$;
    \item Filtering effects in the data processing, that affect SNR linearly.
\end{enumerate}

The phase $\phaseDiff$ in the time interval $\Delta \tTrunc$ is
\begin{equation}
\phaseDiff = 2 \pi \detuning(\PhiQ) \Delta \tTrunc,
\end{equation}
while the noise of the phase measurement is
\begin{equation}
\sigma_{\phaseDiff} = \frac{\sqrt{2} \sigma}{x_0},
\end{equation}
leading to
\begin{equation}
    \SNR = \aSNR' \times \frac{2 \pi \times \detuning(\PhiQ) \times \Delta \tTrunc \exp\left(-(t/\Ttwostar(\PhiQ)^\alpha)\right)}{\sqrt{2} \sigma}.
\end{equation}
Here, $\aSNR'$ is a constant that accounts for filtering effects in data processing.
As we cannot distinguish between $\sigma$ and effects of filtering $\aSNR'$, and $\aSNR'$ is unknown, we can absorb all multiplicative factors
\begin{equation}
    \SNR = \aSNR'' \times\detuning(\PhiQ) \exp\left(-(\tTrunc/\Ttwostar(\PhiQ))^\alpha\right).
\end{equation}

To evaluate the model, we use a quadratic dependence of the detuning on the flux $\detuning(\PhiQ) = a\PhiQ^2$. We find that this dependence matches well the experimentally measured dependence in studied range of pulse amplitudes, up to 0.33 $\Phi_0$ for $a=16.9$~$\frac{\mathrm{GHz}}{\Phi_0^2}$ (\cref{fig:DAC_arc}). Furthermore, we use $\alpha=1$ and a dephasing rate given by
\begin{equation}
    \frac{1}{\Ttwostar(\PhiQ)} = \Gamma = \Gamma_0 + \Gamma_1 \left\lvert\frac{\dd \detuning}{\dd \PhiQ}\right\rvert,
\end{equation}
where $\Gamma_0$ describes the flux-independent dephasing, and $\Gamma_1$ parametrizes the contribution to dephasing due to $1/f$ flux noise~\cite{Martinis03}.
Finally, in the limit of small errors:
\begin{equation}
    \SNR = \aSNR \PhiQ^2 \exp \left( -(\Gamma_0 + 2 a \Gamma_1 \PhiQ) t \right).
\end{equation}
% \cref{fig:SNR}
The joint fit to all data in Fig. 3 uses $\Gamma_0 = 66.7\cdot 10^{-3} ~\s^{-1}$, corresponding to the measured sweetspot $T_2^* = 15$~$\mu$s and yields ${\Gamma_1 = 0.213 \times 10^{-3}} \ \Phio$ equivalent to $\sqrt{A_\Phi} = 12 \times 10^{-6} \ \Phio$ for the single-sided flux noise power spectrum $S(f) = A_\Phi/f$ in a reasonable agreement with the typically reported values\cite{Bylander11,Anton13,Kumar16,Quintana17} and consistent with values measured in our group~\cite{Luthi18, Rol19_NetZero_S}.

\begin{figure}
  \centering
  \includegraphics[width=0.45\textwidth]{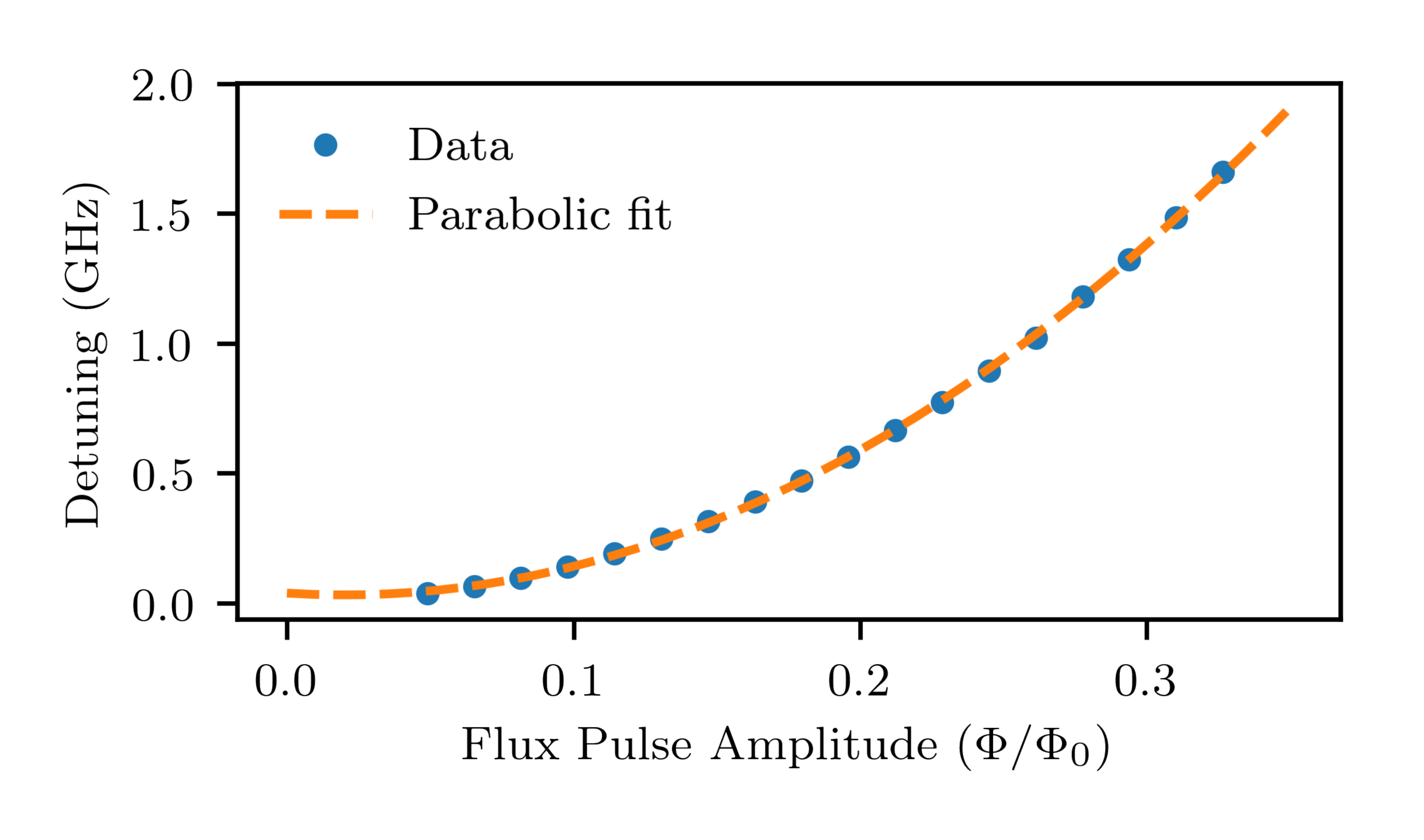}
  \caption{\label{fig:DAC_arc}
 Measured dependence of detuning on the applied flux through the SQUID loop of the studied transmon qubit.
  }
\end{figure}

\section{Real-time predistortion filters}
We make use of two types of digital filters to correct for distortions in real time: a finite impulse response (FIR) filter for short-timescale ($<30~\ns$) distortions and a first-order infinite impulse response (IIR) filter.

\subsection{FIR filter}

A FIR filter implements a convolution with the impulse response $\hfir[n]=b_n$
 specified by the coefficients $b_i$ and described by:
\begin{equation}
y[n] = \sum_{i=0}^{N}b_ix[n-i],
\label{eq:filter_generic_FIR}
\end{equation}
where $x[n]$ is the signal at time $n$ at the input of the filter and $y[n]$ is the signal at the output of the filter.

The real-time FIR filter allows specifying 40 parameters to determine the first 72 coefficients $b_i$.
The first 8 parameters directly correspond to the first 8 $b_i$ coefficients while the remaining 32 parameters set pairs of parameters.
Allowing a total of 72 coefficients $b_i$ to be set, corresponding to a filter length of $30~\ns$ for the AWG sampling rate of 2.4 GSa/s.

\subsection{Exponential over- and under-shoot correction IIR filter}
An IIR filter is slightly more complicated than a FIR filter because it includes feedback:
\begin{equation}
a_0 y[n] = \sum_{i=0}^{N}b_ix[n-i] - \sum_{i=1}^{M}a_i y[n-i].
\label{eq:filter_generic_discrete}
\end{equation}
Here, $a_i$ are the feedback coefficients that describe how $y[n]$ depends on values at the output of the filter at preceding times.
\Cref{eq:filter_generic_discrete} is known as the time-domain difference equation and is a generalization of \cref{eq:filter_generic_FIR}.

A first-order IIR filter is implemented in hardware and intended to correct an exponential over- or undershoot in the step response.
For a step response described by
\begin{equation}
\step = g(1+Ae^{-t/\tIIR})\cdot\heaviside,
\label{eq:exp_step_response}
\end{equation}
where $\tIIR$ is the time constant, $A$ the amplitude and $g$ is a gain correction factor that is ignored.
The filter that corrects for this effect is described by the coefficients

\[
b_0 = 1 - k + k\cdot \alpha, \quad
b_1 = -(1 - k)\cdot(1 - \alpha),\\\\
a_0 = 1 \quad \mathrm{and} \quad
a_1 = -(1-\alpha),
\]
with
\[
\alpha= 1- e^{1/{f_s\tIIR(1+A)}}
\quad \textrm{and} \quad
    k =
\begin{cases}
    \frac{A}{(1+A)(1-\alpha)},& \text{if } A < 0\\
    \frac{A}{(1+A-\alpha)},              & \text{if } A\geq 0,
\end{cases}
\]
where $f_s$ is the sampling rate and $a_i, b_i = 0$ for $i>1$.

The limitations of the hardware implementation of the IIR filter can best be described using an equivalent representation of the filter.
The ideal IIR filter obeys the following difference equation mapping the input samples $x[n]$ to the output samples $y[n]$
\begin{equation}
\label{eq:diff_eq_IIR}
y[n] = (1-k)\:x[n]+k\:u[n],
\end{equation}
where $u[n]$ represents the state of the IIR filter, which is determined by the recursive difference equation, known as an exponential moving average
\begin{equation}
\label{eq:exp_moving_avg_filter}
u[n] = u[n-1]+\alpha (x[n]-u[n-1]).
\end{equation}
Implementing the recursion in \cref{eq:exp_moving_avg_filter} directly with state-of-the-art digital signal processing hardware is infeasible due to the high sampling rate (2.4~GSa/s).
Instead, the real-time filters compute the state variable $u[n]$ based on an average of 16 samples.
Furthermore, the IIR filter is operated at a clock frequency of $300~\MHz$, which means that the state variable $u[n]$ gets updated only every 8-th sample.
The down sampled $u[n]$ is combined with the input signal $x[n]$ in \cref{eq:diff_eq_IIR} at the full sampling rate.
These hardware approximations where taken into account when modeling the impulse response of the filter.

\subsection{Modelling filters}
The Python library SciPy~\cite{Scipy01} provides a function called
``lfilter(b, a, sig)'', which applies the filter defined by the coefficient vectors ``b'' and ``a'' to the signal defined by the vector ``sig''.
In this work, we use this function to predict the effect of applying the real-time predistortion filters.
Because the implementations of the real-time digital filters in hardware requires certain modifications to approximate the ideal filter operation there are slight differences between  the real-time filters and the ideal filters.
These deviations are taken into account when we predict the effect of applying a specific filter.

\section{Using Cryoscope to measure arbitrary shapes}
Cryoscope is capable of sampling arbitrary flux control pulses.
To demonstrate this capability we have chosen a typical Amsterdam canal skyline as an example of an arbitrary shape for the flux pulse.
\Cref{fig:amsterdam} demonstrates this capability by providing a near perfect reconstruction of the target waveform.
Note that the reconstruction involves no free parameters.

\begin{figure}
  \centering
    \includegraphics[width=0.45\textwidth]{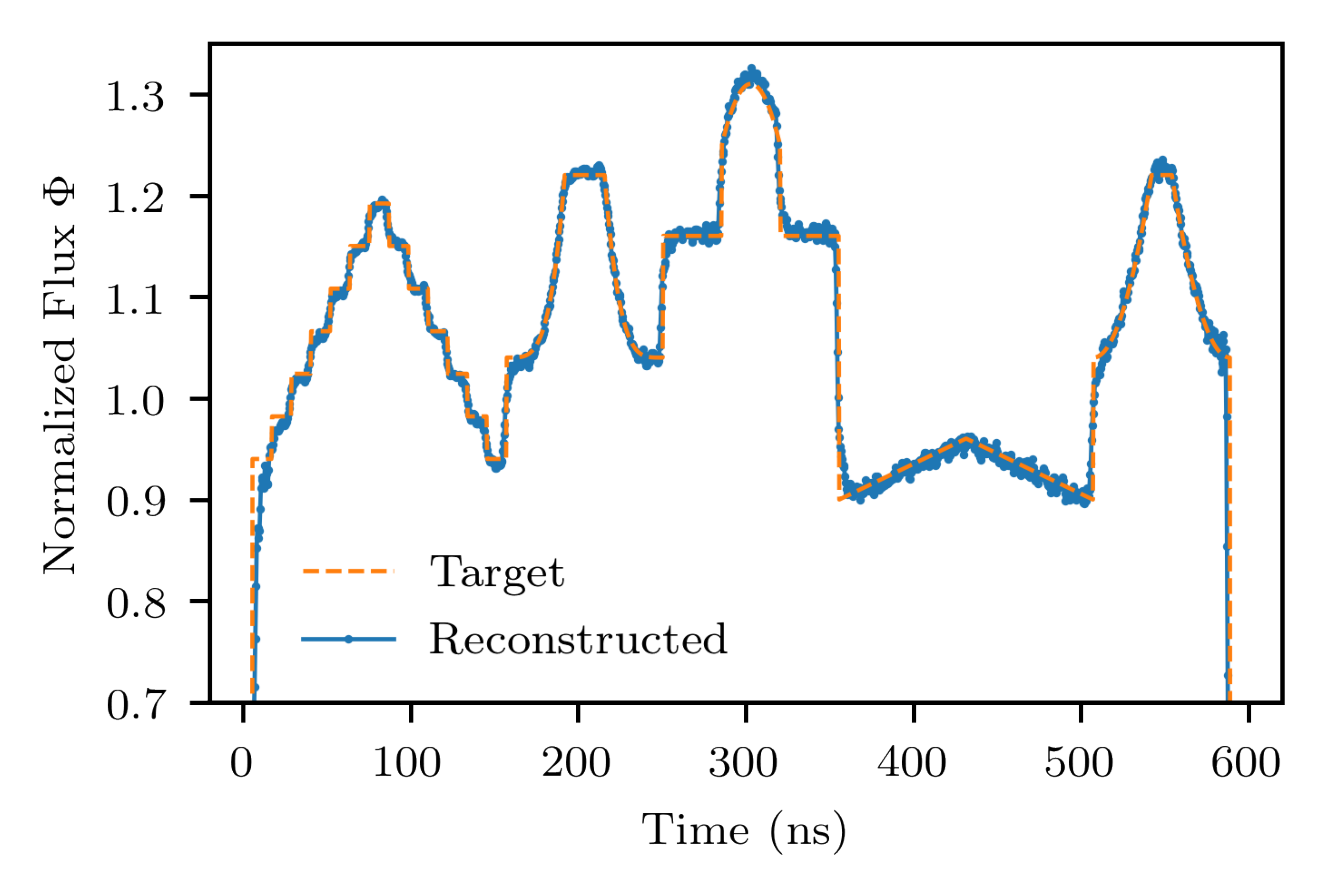}
  \caption{\label{fig:amsterdam}
  Reconstruction of an arbitrary waveform consisting of a typical Amsterdam canal skyline.
  }
\end{figure}
\bibliographystyle{apsrev4-1}
% \bibliography{../../../../Paper_resources/References/References_cQED}
%merlin.mbs apsrev4-1.bst 2010-07-25 4.21a (PWD, AO, DPC) hacked
%Control: key (0)
%Control: author (72) initials jnrlst
%Control: editor formatted (1) identically to author
%Control: production of article title (-1) disabled
%Control: page (0) single
%Control: year (1) truncated
%Control: production of eprint (0) enabled
%

\end{document}